\documentclass[aps,prl,twocolumn,showpacs,superscriptaddress,groupedaddress]{revtex4}  
\usepackage[utf8]{inputenc}
\usepackage{array, tabularx, booktabs}
\usepackage{braket}
\usepackage[english]{babel}
\usepackage{graphicx}
\usepackage{dcolumn}
\usepackage{bm}%
\usepackage{mathtools,lipsum}
\usepackage{mathrsfs}
\usepackage{upgreek}

\usepackage{amsmath} \usepackage{amsfonts} \usepackage{amssymb}
\usepackage{wasysym}
\usepackage[dvipsnames]{xcolor}
\usepackage{graphicx}
\usepackage{multirow}
\usepackage{here}
\usepackage{epsfig}
\usepackage{epstopdf}
\usepackage{soul}
\usepackage{hyperref}
\usepackage{float}
\usepackage{xcolor}
\usepackage{graphicx}
\usepackage{multirow}
\usepackage{ulem}
\usepackage{array}
\usepackage{cancel}
\newcolumntype{L}{>{\displaystyle}l}
\newcolumntype{C}{>{\displaystyle}c}
\newcolumntype{R}{>{\displaystyle}r}
\newcommand{\be}{\begin{equation}}
\newcommand{\ee}{\end{equation}}
\newcommand{\bea}{\begin{eqnarray}}
\newcommand{\eea}{\end{eqnarray}}

\begin{document}
\title{The Quantum Focusing Conjecture and the Improved Energy Condition}
\author{Ido Ben-Dayan}
\affiliation{Physics Department, Ariel University, Ariel 40700, Israel 
}

\date{\today}

\email{ido.bendayan@gmail.com}

\begin{abstract}
 By rearranging its terms, the Quantum Focusing Conjecture (QFC) can be viewed as a quantum energy condition, and we can consider various limits. A recent restricted version is a limiting form where the quantum focusing vanishes $\Theta \rightarrow 0$, and has been proven for Braneworld scenario. As a result, we derive an improved quantum null energy condition (INEC) $T_{kk}\geq \frac{\hbar}{2\pi \mathcal{A}}\left(S''_{out}-\frac{1}{2}\theta S'_{out}\right)$, that can be proven with field theory techniques. We sketch the beginning of a proof, and briefly discuss possible interpretations in the absence of one.   
 
\begin{description}
\item[Dedication]{\centering \textit{In memory of over 1400 dead and in prayer for a speedy release of over 200 hostages of the antisemitic attack on Simchat Torah, October 7th, 2023.} 
\par}
\end{description}
\end{abstract}

\maketitle


\section{Background}
Einstein's theory of General Relativity (GR) is a classical non-linear field theory that connects geometry $g_{\mu \nu}$ with energy $T_{\mu \nu}$ via Einstein's field equations (EFE):
\be
G_{\mu \nu}=8\pi G T_{\mu \nu}.
\ee
In the absence of additional input, any metric ansatz $g_{\mu \nu}$ is allowed. As a result, one can have arbitrary forms of energy by simply postulating an ansatz $g_{\mu \nu}$ and deriving the energy-momentum tensor using EFE. This is the so called Synge's method. Given that experimentally we do not observe arbitrary forms of energy, and for the sake of predictivity, we generally add this additional input in the form of some plausible energy condition on top of EFE. For example, we say that the energy content needs to fulfill the inequality $T_{\mu \nu}k^{\mu}k^{\nu}\geq 0$, where $k^{\mu}$ is any null future pointing vector. This is the celebrated Null Energy Condition (NEC). The NEC and other energy conditions were very instrumental in deriving singularity theorems in GR, usually via the Raychaudhuri equation in $D$ dimensions,
\be
\frac{d\theta}{d\lambda}=-\frac{1}{D-2}\theta^2-\sigma^2-R_{kk}
\ee
and substituting $R_{kk}\equiv R_{\mu \nu}k^{\mu}k^{\nu}=8\pi G T_{ab}k^{\mu}k^{\nu}$ from EFE.
Here $\theta\equiv \frac{1}{\mathcal{A}}\frac{d \mathcal{A}}{d \lambda}$ is the expansion scalar, $\sigma^2$ is the shear term, $\lambda$ is the affine parameter, and $\mathcal{A}$ is an infinitesimal area element spanned by nearby null geodesics. Assuming the NEC means that
\be
\frac{d\theta}{d\lambda}=\frac{d}{d\lambda}\left(\frac{1}{\mathcal{A}}\frac{d \mathcal{A}}{d \lambda}\right)\leq 0.
\ee
Hence a caustic is reached at finite affine parameter, and with further analysis, existence of a singularity in spacetime can be proven \cite{Wald:1984rg}.
However, problems arise once we consider quantum fields which are the basis of contemporary physics. First, for large enough energy densities $T_{\mu \nu}\gtrsim G^{-2}$ quantum fluctuations are so strong that it backreacts on the geometry such that the geometry is not well-defined, requiring a theory of quantum gravity. Second, even in the semiclassical approximation, $T_{\mu \nu}\ll G^{-2}$ quantum fields are known to violate the NEC. Since semiclassically singularities still occur, it is highly desired to find energy conditions that are valid also for quantum fields, preferably based on field theory in flat space, such that they will be valid also in modified and even quantum gravity. One such example is the Quantum Null Energy Condition (QNEC) \cite{Bousso:2015mna, Bousso:2015wca}. 

Investigation of black holes, their thermodynamics and their quantum nature are a major effort in understanding quantum gravity and its semiclassical approximation. A major success is the notion of generalized entropy $S_{gen}$ and the generalized second law of thermodynamics (GSL)
\bea
S_{gen}=\frac{k_Bc^3}{4G\hbar}A+S_{out},\\
dS_{gen}\geq 0,
\eea
which were originally proposed as a way to reconcile thermodynamics and black holes, where $A$ is the area of the black hole, and $S_{out}$ was the ordinary entropy of matter outside the black hole horizon. The profound nature of the generalized entropy is evident from the fact that all the fundamental constants of Nature appear in it. 

By now $S_{gen}$ has been generalized to more general settings of infinitesimal area $\mathcal{A}$ of a hypersurface and $S_{out}$ is the entanglement entropy of the quantum state across the hypersurface. The generalized entropy is expected to be finite, and is of mixed nature as the area term is local, while $S_{out}$ is inherently non-local. Nevertheless, significant progress has been achieved in investigating $S_{gen}$ and its consequences for semiclassical and quantum gravity. A recent very successful program is considering classical GR results, and replacing $\mathcal{A}\rightarrow 4G\hbar S_{gen}$ (From now on $k_B=c=1$), e.g. \cite{Bousso:2018bli,Bousso:2022ntt} and references therein.

\begin{figure}
    \centering
    \includegraphics[scale=0.3]{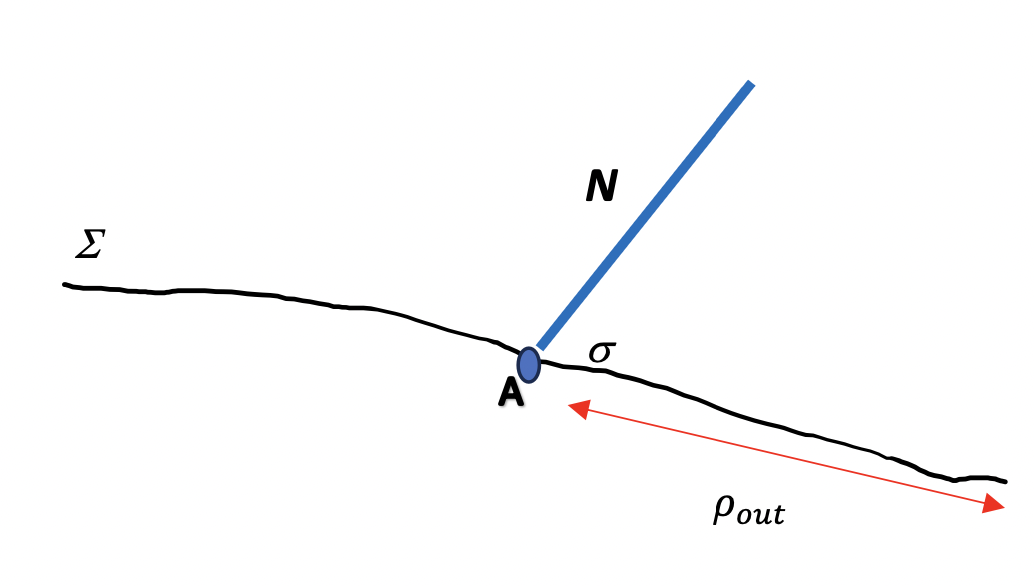}
\includegraphics[scale=0.3]{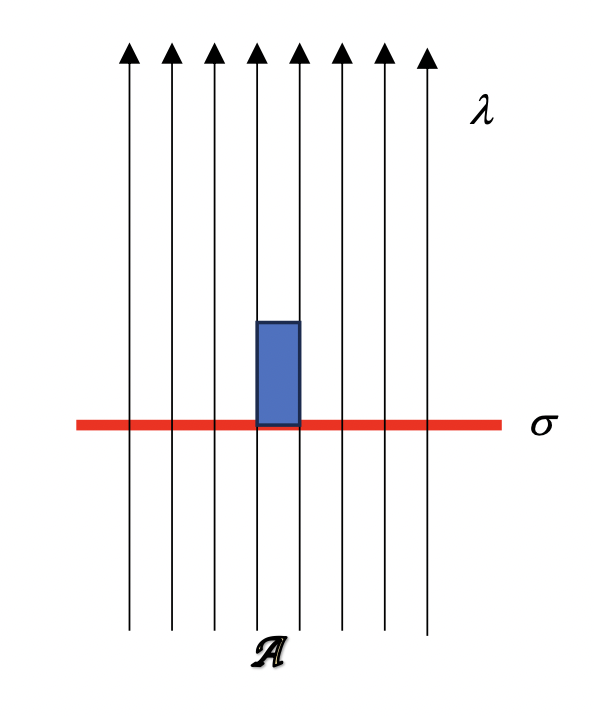}
    \caption{Upper panel: $\Sigma$ is a Cauchy surface that $\sigma$ splits into two parts. $N$ is a null hypersurface and $\rho_{out}$ is the reduced density matrix to one side of $\sigma$. $A$ is the area of $\sigma$. Lower panel: $N$ is divided into pencils of width $\mathcal{A}$ around its null generators. $\sigma$ is deformed along the null generators shown as the blue rectangle. Since the off-diagonal QFC has already been proven, we will be interested solely in variations of the diagonal type.}
   \label{fig:1dpicture}
\end{figure}

Within this program, a quantum expansion scalar $\Theta$, has been defined \cite{Bousso:2015mna}. Consider a spatial surface $\sigma$ which is a codimension-2 hypersurface of area $A$ that splits a Cauchy surface $\Sigma$ in two. $S_{out}$ is the entanglement entropy of the quantum state on one side of $\sigma$. $N$ is an orthogonal null hypersurface (codimension-1), which is divided
into pencils of width $\mathcal{A}$ around its null generators. $\sigma$ is then deformed by an affine parameter. The quantum expansion $\Theta$ is the response of $S_{gen}=\frac{A}{4G\hbar}+S_{out}$ to deformations of $\sigma$ along $N$:
\be
\Theta[V(y),y_1]\equiv\frac{4 G \hbar}{\sqrt{V_g(y_1)}}\frac{\delta S_{gen}}{\delta V(y_1)}
\ee
where $V_g$ is the (finite) area element of the metric restricted to $\sigma$. The notation $\Theta[V (y_1), y_1]$ emphasizes that the quantum expansion requires the
specification of a slice $V (y_1)$ and is a function of the coordinate $y_1$ on that slice. An example of its use is the Quantum Extremal Surface (QES), a surface where $\Theta=0$, \cite{Engelhardt:2014gca}, which has been crucial in recent developments regarding the black hole information paradox \cite{Almheiri:2019psf,Penington:2019npb}.
With the advent of the quantum expansion scalar, a Quantum Focusing Conjecture (QFC) has been proposed \cite{Bousso:2015mna}:
\be
\frac{\delta}{\delta V(y_2)}\Theta[V(y_1),y_1]\leq 0
\ee
In words, the quantum expansion cannot increase at $y_1$, if the slice of $N$ defined by $V (y_1)$
is infinitesimally deformed along the generator $y_2$ of $N$, in the same direction. Here $y_2$
can be taken to be either the same or different from $y_1$. The pictorial setup, is drawn in figures \ref{fig:1dpicture},\ref{fig:2dpicture}. The non-diagonal part of $y_1\neq y_2$ has been proven using entropy subadditivity when the QFC was proposed in \cite{Bousso:2015mna}. The diagonal part $y_1=y_2$ is still a conjecture.
Equivalently, we can write the diagonal part as a more pedestrian expression:
\be \label{eq:Theta}
\Theta\equiv\theta+\frac{4G\hbar}{\mathcal{A}}S_{out}',
\ee
where a prime denotes a derivative with respect to the affine parameter $\lambda$.
So far, many results were derived using the QFC, no counter examples have been found, but a proof still evades the scientific community. We will show that the QFC implies a new stronger energy condition and sketch a possible attempt to prove it.

\section{The diagonal part of the QFC in the $\Theta\rightarrow 0$ limit}
From
\be \label{eq:Theta}
\Theta=\theta+\frac{4G\hbar}{\mathcal{A}}S_{out}',
\ee
the diagonal part of the QFC is the statement that
\be
\theta'+\frac{4G\hbar}{\mathcal{A}}(S_{out}''-\theta S_{out}')=\Theta' \leq 0
\ee
Using the Raychaudhuri equation, the EFE, and zero shear $\sigma^2=0$, implies
\be
\Theta'=-\frac{1}{D-2}\theta^2 -8\pi G T_{kk}+\frac{4G\hbar}{\mathcal{A}}(S_{out}''-\theta S_{out}')\leq 0
\ee
where $T_{kk}\equiv \langle T_{\mu \nu} k^{\mu} k^{\nu}\rangle$ is the  quantum corrected expectation value of the energy-momentum tensor contracted with null vectors.
Thus, the QFC is basically a quantum energy condition in the presence of gravity:
\be
T_{kk}\geq \frac{\hbar}{2\pi \mathcal{A}}\left(S''_{out}-\theta S'_{out}\right)-\frac{1}{(D-2)8\pi G}\theta^2 \label{eq:GNEC}
\ee
If this conjecture is to hold always, then it should certainly hold in limiting cases. 
The limit of $\hbar\rightarrow 0$ simply reproduces the aforementioned classical NEC: 
\be
T_{kk}\geq 0.
\ee 
However, it is known that quantum effects where $\hbar \neq 0$, violate this condition.
Next, one can consider the limiting case where the classical expansion vanishes, $\theta\rightarrow 0$, which results in the QNEC:
\be
T_{kk}\geq \frac{\hbar}{2\pi \mathcal{A}}S''_{out}
\ee
This inequality has been proven in certain cases \cite{Bousso:2015wca,Koeller:2015qmn,Balakrishnan:2017bjg,Malik:2019dpg,Moosa:2020jwt,Roy:2022yzm}. 
There are also examples where this inequality is saturated, under the assumption of $\theta=0$, \cite{Balakrishnan:2019gxl,Leichenauer:2018obf}.
Nevertheless, one can consider other limiting cases \footnote{Presumably, the vanishing of $S''_{out}$ or $S'_{out}$ implies that the bound on $T_{kk}$ is lower. However, such limits are trivially satisfied if the QNEC or the soon to be discussed INEC hold.}.

For two reasons, perhaps the most interesting limit is when the \textit{quantum expansion} vanishes, $\Theta\rightarrow 0$.
The first reason is the Quantum Extremal Surface, $\Theta=0$ and its applications. The second reason, is a recently proposed restricted version of the QFC  \cite{Shahbazi-Moghaddam:2022hbw}, that wherever $\Theta=0$ pointwise, then $\Theta'\leq 0$. This restricted version has been proven in Braneworld scenario. As such, it is much more well established. This is a major breakthrough, especially since many of the interesting results derived using the original QFC still hold considering the restricted version \cite{Brown:2019rox,Akers:2019lzs,Engelhardt:2021mue,Akers:2016ugt,Wall:2010jtc,Bousso:2022tdb,Bousso:2015eda,Balakrishnan:2017bjg,Ceyhan:2018zfg}.
Considering $\Theta=0$ implies \footnote{Obviously, the possibility of $\Theta=\theta=S_{out}'=0$ is a private case of this limit. This private case simply reproduces the original QNEC. We are interested in situation where this is not the case.}
\be
\theta=-\frac{4G\hbar}{\mathcal{A}}S'_{out}
\ee
Since the QFC or its restricted version is expected to always hold, we can substitute this equality into \eqref{eq:GNEC}, in 4D \footnote{In any other dimension $D\geq 3$, one simply needs to replace $1/2\rightarrow (D-3)/(D-2)$ in the inequality, and the more concise version is $T_{kk}\geq \frac{\hbar}{2\pi \mathcal{A}^{1/D-2}}\left(\frac{S'_{out}}{\mathcal{A}^{D-3/D-2}}\right)'$}:
\be \label{eq:BEC4D}
T_{kk}\geq \frac{\hbar}{2\pi \mathcal{A}}\left(S''_{out}-\frac{1}{2}\theta S'_{out}\right)=\frac{\hbar}{2\pi \sqrt{\mathcal{A}}}\left(\frac{S'_{out}}{\sqrt{\mathcal{A}}}\right)'.
\ee
Thus, we have derived an "improved" quantum null energy condition (INEC).
Notice, that similarly to the original QNEC, the modified one does not have $G$ in it. It is a field theory statement that could be analyzed using field theory techniques in flat space. Moreover, despite the minus sign, this term is positive definite. Hence, it is a stronger inequality than the QNEC, that will be more difficult to satisfy.
To see that this is a positive definite term, we can reuse the $\Theta=0$ equation to write the INEC just in terms of the derivatives of $S_{out}$:
\be
T_{kk}\geq \frac{\hbar}{2\pi \mathcal{A}}\left(S''_{out}+\frac{2 G\hbar}{\mathcal{A}} S_{out}^{'2}\right)\geq\frac{\hbar}{2\pi \mathcal{A}}S''_{out}.
\ee
For the sake of completeness, one can also rewrite the inequality as a bound on $S''_{out}$ using geometrical quantities:
\be
T_{kk}\geq \frac{\hbar}{2\pi \mathcal{A}}S''_{out}+\frac{1}{16\pi G}\theta^2 \Leftrightarrow 
R_{kk}-\frac{1}{2}\theta^2\geq \frac{4G\hbar}{\mathcal{A}}S''_{out}.
\ee

While the restricted QFC has been proven only in Braneworld scenario, we think the above possibilities are relevant in general for both the QFC and the restricted version. There are four possible interpretations of the results derived here. 
\begin{enumerate}
    \item The INEC is true and can be proven using field theory techniques.
    \item If a counter example to the INEC is found, then both the QFC and its restricted version are false. If not, but we fail to prove the INEC, then both QFC and the restricted version hold only as perturbative statements, with the small parameter being $\frac{G\hbar}{\mathcal{A}}\ll 1$. Thus, the second term $\sim \theta S'_{out}$ is always subdominant and can be neglected, such that it cannot spoil the QNEC. Alternatively, the QFC needs to be modified by adding additional terms with contributions of the order of $\frac{G\hbar}{\mathcal{A}}\ll 1$. These contribution may come from curvature corrections, or corrections to $S_{out}$. 
    \item Points where $\Theta=0$ are realized only in the redundant case of $\Theta=\theta=S'_{out}=0$ and the QNEC is saturated only in these points $T_{kk}=\frac{\hbar}{2\pi \mathcal{A}}S''_{out}$.
    \item Contrary to the standard lore where $T_{kk}$ is viewed as the quantum corrected energy momentum tensor to all orders in $\hbar$, we should think of it as an expansion in $\hbar$ and compare order by order, $T_{kk}=T^{(0)}_{kk}+\hbar T^{(1)}_{kk}+\cdots$, and the Einstein's equations are valid for the classical part $R_{ab}k^ak^b=T^{(0)}_{kk}$. Thus, the classical Raychaudhuri equation removes this term and the remaining inequality is regarding the first order quantum correction:
    \be
    T^{(1)}_{kk}\geq \frac{1}{2\pi \mathcal{A}}(S_{out}''-\theta S'_{out}).
    \ee
    This is a pointwise version of the QHANEC \cite{Akers:2016ugt}, for the first order quantum correction $T^{(1)}_{kk}$, see more below.
     This is the QFC result, the restricted QFC will further imply 
    \be
    T^{(1)}_{kk}\geq \frac{1}{2\pi \mathcal{A}}\left(S_{out}''+\frac{4 G\hbar}{\mathcal{A}} S^{'2}_{out}\right).
    \ee
\end{enumerate}
Finally, considering the INEC in general dimensions, 
\bea \label{eq:BECgeneralD}
T_{kk}\geq \frac{\hbar}{2\pi \mathcal{A}}\left(S''_{out}-\frac{D-3}{D-2}\theta S'_{out}\right)\cr
=\frac{\hbar}{2\pi \mathcal{A}^{1/D-2}}\left(\frac{S'_{out}}{\mathcal{A}^{D-3/D-2}}\right)',
\eea
we find two interesting limits. 
First, $D=3$ is a special case as the terms containing $S'_{out}$ vanish exactly and the INEC reduces to the form of the original QNEC. Thus, in $D=3$ the $\Theta\rightarrow 0$ limit does not yield new insights beyond the original QNEC. Current examples of QNEC saturation $T_{kk}= \frac{\hbar}{2\pi \mathcal{A}}S''_{out}$ always occur on $\theta=0$ hypersurfaces. It may be interesting to investigate whether there are examples in $D=3$ where the saturation occurs even at points where $\theta \neq0$. 

Second, considering the infinite dimension limit $D\rightarrow \infty$: 
\be
T_{kk}\geq \frac{\hbar}{2\pi \mathcal{A}}\left(S''_{out}-\theta S_{out}'\right)=\frac{\hbar}{2\pi}\left(\frac{S_{out}'}{\mathcal{A}}\right)',
\ee
which is a pointwise version of the QHANEC \cite{Akers:2016ugt}. To see this, integrate from the Cauchy splitting hypersurface along the affine parameter to infinity, and assume $\frac{S_{out}'}{\mathcal{A}}$ vanishes at infinity yielding:
\be
\int_\lambda^\infty d\lambda'  \,T_{kk}(\lambda') \geq -\frac{\hbar}{2\pi\mathcal{A}}S_{out}',
\ee
which is the QHANEC.

\section{Sketch of a possible proof}
\begin{figure}
    \centering
    \includegraphics[scale=0.3]{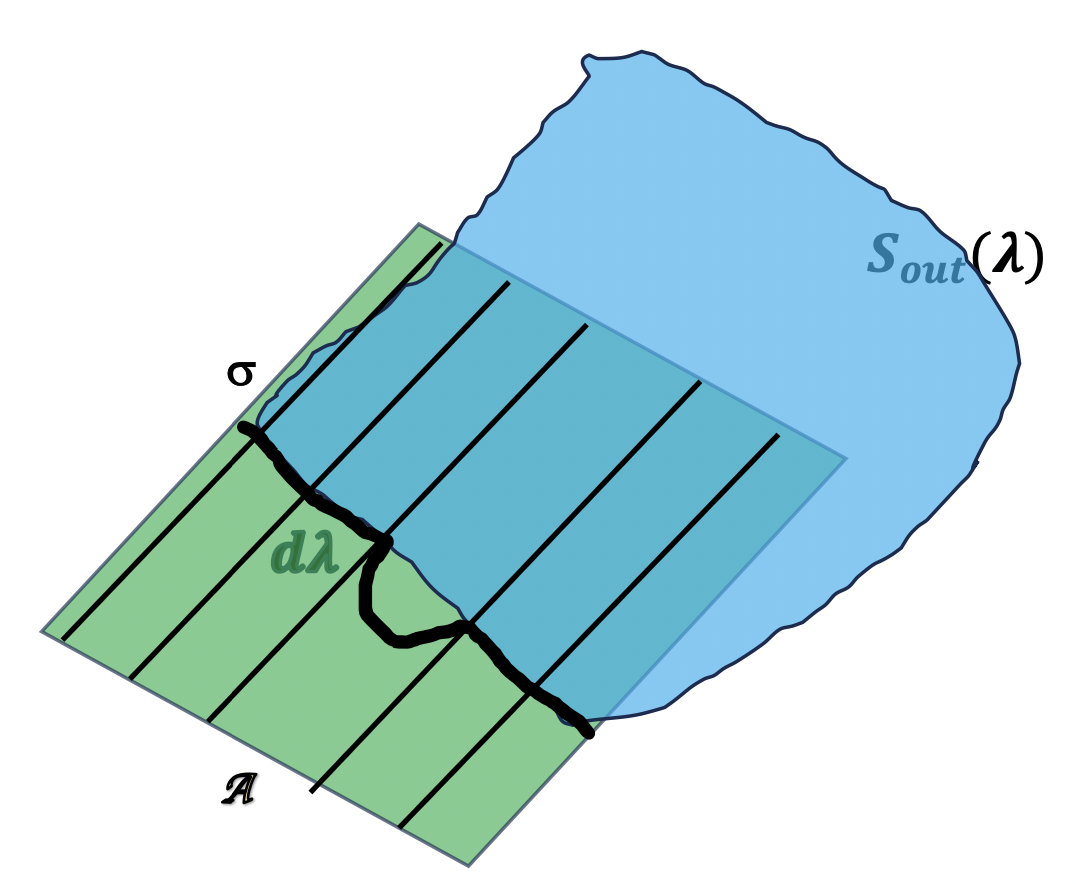}
    \caption{An example of the setup. The quantum expansion $\Theta$ at one point of $\sigma$ is the rate at which the generalized entropy $S_{gen}$ changes under a small variation $d\lambda$ of $\sigma$, per cross-sectional area $\mathcal{A}$ of the variation. According to the QFC, $\Theta'\leq 0$, i.e. $\Theta$ cannot increase under a second variation in the same direction.}
    \label{fig:2dpicture}
\end{figure}
The QNEC has been proven in various settings and different methods by now. We shall follow the method of proof used in \cite{Bousso:2015wca} in stating what has to be evaluated to be considered a proof. The proof relied on null quantization. Recall that $\sigma$ splits the Cauchy surface $\Sigma$ in two, 
and that $N$ is an orthogonal null hypersurface, which is divided
into pencils of width $\mathcal{A}$ around its null generators. $\sigma$ is then deformed by an affine parameter, see figure \ref{fig:2dpicture}. 
For small $\mathcal{A}$ the pencil state should be very close to the vacuum state. We can then write the state as

\begin{equation}\label{eq-state}
\rho(\lambda) = \rho^{(0)}_{\rm pen}(\lambda) \otimes \rho^{(0)}_{\rm aux} + \sigma(\lambda)~,
\end{equation}
where $\rho^{(0)}_{\rm pen}(\lambda)$ is the vacuum state density matrix on the part of the pencil with affine parameter greater than $\lambda$, $\rho^{(0)}_{\rm aux}$ is some state in an auxiliary system, and the perturbation $\sigma(\lambda)$ is small 
The corresponding entanglement entropy of $\rho(\lambda)$ is $S_\text{out}(\lambda)$. We will expand it as a perturbation series in $\sigma(\lambda)$:
\be\label{eq-Slambda}
S_{\rm out}(\lambda) = S^{(0)}(\lambda) + S^{(1)}(\lambda) + S^{(2)}(\lambda) + \cdots
\ee
where $S^{(n)}(\lambda)$ contains $n$ powers of $\sigma(\lambda)$, and we suppress the subscript $out$ from now on. At zeroth order, since $\rho^{(0)}$ is a product state, we have
\begin{eqnarray}
S^{(0)}(\lambda) &=& - Tr\left[\rho^{(0)}(\lambda) \log \rho^{(0)}(\lambda)\right]\cr
&=&  - Tr\left[\rho_{\rm pen}^{(0)}(\lambda) \log \rho_{\rm pen}^{(0)}(\lambda)\right]
- Tr\left[\rho_{\rm aux}^{(0)} \log \rho_{\rm aux}^{(0)}\right].\cr
\end{eqnarray}
The first term on the right-hand side is independent of $\lambda$ because of null translation invariance of the vacuum: all half-pencils have the same vacuum entropy. The second term is manifestly independent of $\lambda$. So $S^{(0)}$ is $\lambda$-independent and will not affect the INEC.
Considering the next term, $S^{(1)}(\lambda)$:
\begin{eqnarray}
S^{(1)}(\lambda) &=& -Tr\left[\sigma(\lambda) \log \rho^{(0)}(\lambda)\right] \cr
&=& -Tr\left[\sigma(\lambda) \log \rho_{\rm pen}^{(0)}(\lambda)\right] -Tr\left[\sigma(\lambda) \log \rho_{\rm aux}^{(0)}\right].\cr
\end{eqnarray}
Let us analyze the second term on the right hand side by tracing over the pencil subsystem:
\bea
Tr\left[\sigma(\lambda) \log \rho_{\rm aux}^{(0)}\right] = Tr_{\rm aux}\left[\left[Tr_{\rm pen}\sigma(\lambda) \right]\log \rho_{\rm aux}^{(0)}\right] \cr
= Tr_{\rm aux}\left[\sigma(\infty)\log \rho_{\rm aux}^{(0)}\right].
\eea
Thus, this term is $\lambda$ independent.
For the first term, the fact that $\rho^{(0)}_{\rm pen}(\lambda)$ is thermal with respect to the boost operator on the pencil implies:
\be
-Tr\left[\sigma(\lambda) \log \rho^{(0)}_{\rm pen}(\lambda)\right] = \frac{2\pi \mathcal{A}}{\hbar} \int_\lambda^\infty d\lambda'  \,(\lambda'-\lambda) T_{kk}(\lambda'),
\ee
where the integral is along the generator which defines the pencil and the expectation value is taken in the excited state. This is the first $\lambda$-dependent term we have in the perturbative expansion of $S(\lambda)$. Taking one and two derivatives and evaluating at $\lambda=0$ gives the following identities:

\begin{widetext}
\begin{eqnarray} \label{eq:Sidentitiesnew}
\left(S^{(0)} + S^{(1)} \right)' &=&-\frac{2\pi \mathcal{A}}{\hbar}\left(\int_\lambda^\infty d\lambda'  \, T_{kk}(\lambda')-\theta \int_\lambda^\infty d\lambda'(\lambda'-\lambda)  \, T_{kk}(\lambda')\right)\cr
\left(S^{(0)} + S^{(1)} \right)'' &=&  \frac{2\pi \mathcal{A}}{\hbar}\left( T_{kk}-2\theta\int_\lambda^\infty d\lambda'  \, T_{kk}(\lambda')+\frac{1}{2}\theta^2\int_\lambda^\infty d\lambda'(\lambda'-\lambda)  \, T_{kk}(\lambda')\right),
\end{eqnarray}
\end{widetext}
where we have used the fact that in flat space $\theta'=-1/2\theta^2$.
Substituting \eqref{eq-Slambda}\eqref{eq:Sidentitiesnew} into \eqref{eq:BEC4D} 
gives
\bea
\frac{3\pi\mathcal{A}}{\hbar}\,\theta \int_\lambda^\infty d\lambda'  \,T_{kk}(\lambda') \geq S^{(2)''}-\frac{1}{2}\theta S^{(2)'}\cr
\Leftrightarrow 3\pi\theta \int_\lambda^\infty d\lambda'  \, T_{kk}(\lambda') \geq \frac{\hbar}{\sqrt{\mathcal{A}}}\left(\frac{S^{(2)'}}{\sqrt{\mathcal{A}}}\right)'
\eea
Again, the $\theta\rightarrow 0$ limit reproduces the standard requirement in one of the proofs of the QNEC $S^{(2)''}\leq0$, \cite{Bousso:2015wca}. The generalization to any dimension is straightforward.
For the completion of the proof one has to use the replica trick and calculate the same sheet and multi-sheet correlators. We shall attempt such a completion in the near future.
\section*{Acknowledgements}
I thank Arvin Shahbazi-Moghaddam for early collaboration on the project and I wish to thank him, Misha Smolkin and Amir Tajdini for useful discussions.

\end{document}